\documentclass[reprint,aps,prd,nofootinbib,a4paper,10pt,twocolumn,superscriptaddress]{revtex4-2}
\usepackage{hyperref} 
\usepackage{changes}
\usepackage{graphicx,subfigure}
\usepackage[section]{placeins}
\usepackage[titletoc]{appendix}
\usepackage{xcolor}
\usepackage{dsfont}
\usepackage{bm}
\usepackage{mathtools} 
\usepackage{amsfonts}
\usepackage{enumerate}

\usepackage{dcolumn}
\usepackage{multirow}
\usepackage{enumitem}

\relax

\hypersetup{
     colorlinks   = true,
     citecolor    = cyan,
     linkcolor = pink
     }

\begin{document}

    \title{A First Observational Assessment of Cosmic Backreaction Over an Extended Redshift Range\\\small \em Addendum to ``Observational Tests for Distinguishing Classes of Cosmological Models'' (arXiv:2604.07244v1)}

\author{S. M. Koksbang} 
\email{koksbang@cp3.sdu.dk}
\affiliation{CP$^3$-Origins, University of Southern Denmark, Campusvej 55, DK-5230 Odense M, Denmark}

\begin{abstract}
In the recent preprint arXiv:2604.07244v1, the authors introduce a novel combination of redshift, distance, and expansion rate observables for constraining cosmic backreaction observationally. The current work presents a first application of the method, in principle yielding the first direct constraints on the total cosmic backreaction in our universe over a significant redshift range. However, we find that current data are not yet sufficient to place tight constraints on backreaction. Thus, neither vanishing nor significant backreaction can be ruled out with the presented constraints. Nonetheless, the results suggest that forthcoming survey data will enable enlightening constraints on cosmic backreaction using this method.
\end{abstract}

\keywords{cosmology -- beyond $\Lambda$CDM, cosmic tensions}
\maketitle

\section{Introduction}
For decades, the $\Lambda$CDM model has provided an extraordinarily successful description of the Universe. However, in recent years, a growing body of high‑precision observational data has revealed statistically significant tensions that challenge the internal consistency of this model. Among the most prominent tensions is the Hubble tension \cite{hubble_tension} which refers to the persistent discrepancy between early‑ and late‑time measurements of the present‑day expansion rate, $H_0$. Another tension that has attracted considerable attention over the past two years arises from results of the Dark Energy Spectroscopic Instrument (DESI). These results suggest that dark energy may be dynamical rather than a simple cosmological constant \cite{DESI_DE, DESI_DE2, DESI_DE3}. In addition, increasingly robust evidence has emerged for large‑scale anisotropies that appear inconsistent with $\Lambda$CDM expectations when compared across different cosmological probes \cite{dipole1, dipole2, anisotropy_marco}.
\newline\indent
These developments have motivated renewed interest in cosmological models that extend beyond the Friedmann–Lemaitre–Robertson–Walker (FLRW) class of solutions to Einstein’s field equations. In this context, the recent arXiv preprint \cite{preprint} introduces observational methods for constraining parameters within the Dyer–Roeder framework \cite{DR} as well as for quantifying cosmic backreaction \cite{buchert1,buchert2}.
\newline\indent
In the present work, we will apply this approach to current data. For completeness, and to keep the work self-contained, we begin with a brief introduction to the concept of cosmic backreaction. We then restate (and slightly reformulate) the key result of \cite{preprint} relevant here, before presenting the resulting constraints.

\section{Cosmic backreaction}
It is well-known that the large-scale structures of the Universe can affect its average dynamics. This effect is known as cosmic backreaction and is mainly studied by using the Buchert averaging scheme presented in \cite{buchert1, buchert2}. With this scheme, scalar averaging of the Raychaudhuri equation and Hamiltonian constraint is used to obtain dynamical equations for the average/very-large-scale universe. Scalar averaging is defined through the volume-weighted average of a scalar, i.e. 
\begin{align}
    x_\mathcal{D} :=\frac{\int_\mathcal{D} xdV}{\int_\mathcal{D}dV}
\end{align}
for some scalar $x$. Here, $\mathcal{D}$ denotes the spatial domain of averaging which should be above the (expected/assumed) homogeneity scale, while $dV=\sqrt{|{\rm det}(g)|}d^3 x$ is the infinitesimal proper volume element, with $g$ the spatial part of the metric. For details on the setup, the reader is referred to \cite{buchert1}.
\newline\indent
The resulting dynamical equations for the Universe (the averaged Hamiltonian constraint and Raychaudhuri equation) read
\begin{align}\label{eq:friedmann}
    H_\mathcal{D}^2 = \frac{8\pi G}{3}\rho_\mathcal{D} + \Lambda -\frac{R_\mathcal{D} + Q_\mathcal{D}}{6}
\end{align}
and
\begin{align}
    \frac{\ddot a_\mathcal{D}}{a_\mathcal{D}} = -\frac{4\pi G}{3}\rho_\mathcal{D} + \Lambda + Q_\mathcal{D},
\end{align}
where $H_\mathcal{D} = 1/3\cdot \theta_\mathcal{D} = \dot a_\mathcal{D}/a_\mathcal{D}$ is the average of the local expansion rate $\theta$, and $a_\mathcal{D}$ is a volume scale factor defined as the cube root of the proper volume of $\mathcal{D}$, usually normalized to 1 at present time. Although we have neglected pressure in the above, it is straightforward to add this \cite{buchert2}.
\newline\indent
These dynamical equations deviate from those of the FLRW models in two ways. First, there is the genuinely new term
\begin{align}
Q_\mathcal{D}:=2/3\cdot\left((\theta^2)_\mathcal{D}) - (\theta_\mathcal{D})^2\right)-2(\sigma^2)_\mathcal{D},     
\end{align}
where $\sigma^2$ is the shear scalar of the fluid. Second, the curvature term, $R_\mathcal{D}$, will in general behave differently than the FLRW curvature parameter which necessarily scales as $\propto 1/a_\mathcal{D}^2$. Together, these deviations from FLRW dynamics is known as cosmic backreaction.
\newline\indent
In order to more easily assess the importance of the backreaction terms we can define the density parameters 
\begin{align}
    \begin{split}
 \Omega_{R}&:=-R_\mathcal{D}/(6H_\mathcal{D}^2)       \\
 \Omega_Q&:=-Q_\mathcal{D}/(6H_\mathcal{D}^2)\\
 \Omega_m&:=8\pi G\rho_\mathcal{D}/(3H_\mathcal{D}^2)\\
 \Omega_\Lambda&:=\Lambda/(3H_\mathcal{D}^2).
    \end{split}
\end{align}
Introducing these into equation \ref{eq:friedmann}, this equation becomes
\begin{align}\label{eq:one}
    \Omega_m + \Omega_\Lambda + \Omega_R + \Omega_Q = 1.
\end{align}
If the averages are calculated on spatial hypersurfaces of statistical homogeneity and isotropy (assumed to be traced fairly by light), the redshift can, up to statistical fluctuations, be related to $a_\mathcal{D}$ simply through \cite{syksy1, syksy2, selv1, selv2}
\begin{align}\label{eq:z}
    \langle z \rangle = 1/a_\mathcal{D}-1,
\end{align}
where triangular brackets are used to denote averaging over many lines of sight to remove statistical fluctuations. Under these requirements for the averaging hypersurface, the redshift-distance relation can, up to statistical fluctuations, be related to spatial averages through the relation \cite{syksy1, syksy2, selv1, selv2}
\begin{align}\label{eq:DA}
    H_\mathcal{D}\frac{d}{d\langle z \rangle}\left[(1+\langle z\rangle )^2H_\mathcal{D} \frac{d\langle D_A \rangle}{d\langle z\rangle}   \right] = 
    -4\pi G\rho_\mathcal{D}\langle D_A\rangle,
\end{align}
where $D_A$ is the angular diameter distance. Using this, it was in \cite{preprint} concluded that (their equation 8+9)
\begin{align}
    D'' = -\frac{1}{H_\mathcal{D}}H_\mathcal{D}'D' - D\frac{1/6\cdot R_\mathcal{D} + 1/2\cdot Q_\mathcal{D}}{H_\mathcal{D}^2(1+\langle z\rangle)^2},
\end{align}
where primes denote differentiation with respect to $\langle z\rangle$, and where  $D:= (1+\langle z\rangle )\langle D_A\rangle$ (which is not {\em a priori} equal to $\langle (1+z)D_A\rangle $). By rearranging this and introducing the density parameters defined above, we see that we can constrain  the combination $\Omega_R + 3\Omega_Q$ with $D$, $H_\mathcal{D}$ and their derivatives through the relation
\begin{align}\label{eq:test}
    \Omega_R + 3\Omega_Q = \frac{(1+\langle z\rangle)^2}{D}\left[ H_\mathcal{D}'D'/H_\mathcal{D} + D'' \right].
\end{align}
Note that this quantity is slightly different from the quantity $\mathcal{A}:= H_\mathcal{D}^2D''/D + H_\mathcal{D}H_\mathcal{D}'D'/D$ introduced in \cite{preprint}. We chose the slightly different combination here since a constraint directly on the density parameters rather than $R_\mathcal{D}$ and $Q_\mathcal{D}$ themselves is easier to interpret due to the simple relation in equation \ref{eq:one}.
\newline\indent
It is worth emphasizing that the combination on the right hand side reduces to the left hand side {\em only} if spacetime is traced fairly by light. If, on the other hand, light undersamples the density field in the universe e.g. due to opaque regions, the redshift-distance relation of \cite{syksy1, syksy2} is no longer valid. Instead, the differential equation relating the redshift, expansion rate and angular diameter distance with each other will change and the Dyer-Roeder approximation may become more appropriate to use \cite{DR_bc, DR_bc2}, but as emphasized by the last example in \cite{DR_bc2}, the range of validity of the Dyer-Roeder approximation is unclear (see e.g. also \cite{syksy}, especially the end of section 2.3.2). Nonetheless, as seen by the results in \cite{preprint}, the combination on the right hand side of equation \ref{eq:test} can, when assuming the Dyer-Roeder approximation is applicable, instead be used to constrain the $\alpha$ parameter which quantifies the ``unfairness'' of the light's spacetime sampling in the Dyer-Roeder approximation. Specifically, equation \ref{eq:test} would instead be (assuming a dust+$\Lambda$+curvature FLRW universe)
\begin{align}\label{eq:testDR}
\begin{split}
  &\frac{K(1+\langle z\rangle)^2}{H_\mathcal{D}^2} - \frac{3}{2}(1-\alpha)\Omega_m   \\= &\frac{(1+\langle z\rangle )^2}{D}\left[ H_\mathcal{D}'D'/H_\mathcal{D} + D'' \right].
\end{split}
\end{align}
Lastly, there is of course also the option that both backreaction and the Dyer-Roeder approximation are relevant.
\newline\newline
It is unclear to what extent the Dyer-Roeder approximation is relevant for observations made in our universe, i.e. it is unclear to what extent light traces our spacetime fairly. However, as  discussed in e.g. \cite{misinterp}, we expect that light beams of different width will be affected differently by any significant unfair sampling, such that narrow light beams from e.g. supernovae would yield distances in discordance with those obtained from e.g. BAO and CMB which cover the sky or large portions thereof. This would again appear as a violation of the distance-duality relation \cite{etherington} ($D_L = (1+z)^2D_A$, where $D_L$ is the luminosity distance). This relation has been tested abundantly within the last few years, with results being consistent with no violation within $2\sigma$ (see \cite{example1, example2, example3} for a few recent examples using supernovae and BAO data). Thus, we will here assume that any Dyer-Roeder effect can be neglected for the considered observations, i.e. we will assume that $\alpha \approx 1$ and thus that we are indeed constraining $\Omega_R + 3\Omega_Q$.
\newline\indent
Ideally, we wish to constrain the two backreaction parameters $\Omega_R$ and $\Omega_Q$ independently which we see is not possible with the current results of \cite{preprint}. Nonetheless, being able to constrain the combination $\Omega_R + 3\Omega_Q$ is very valuable since both contributions ($\Omega_R$ and $\Omega_Q$) vanish identically in the standard $\Lambda$CDM model.

\section{Constraints}
\begin{figure*}
    \centering
    \includegraphics[width=0.65\columnwidth]{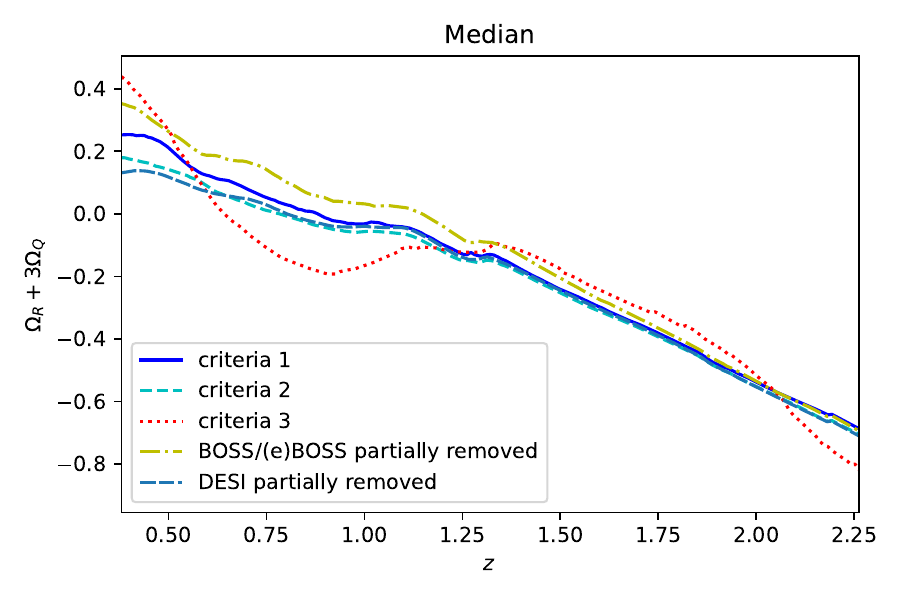}
    \includegraphics[width=0.65\columnwidth]{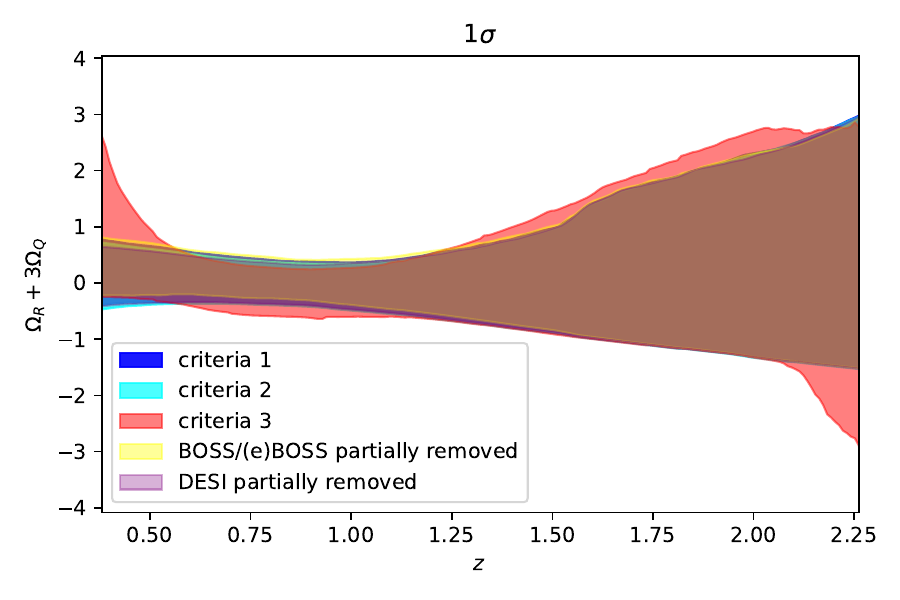}
    \includegraphics[width=0.65\columnwidth]{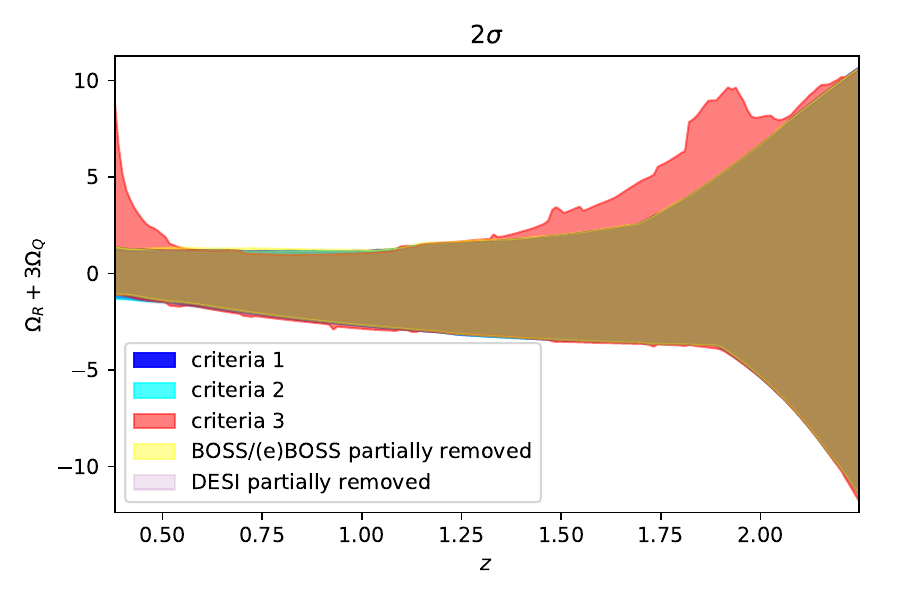}
    \caption{Comparison of median, $1\sigma$ and $2\sigma$ for $\Omega_R + 3\Omega_Q$. The uncertainty bands are almost indistinguishable except for the criteria 3 version. The results were obtained from reconstructing $\langle D_A\rangle $ and $H_\mathcal{D}$ as described in \cite{PRD}. The redshift $z$ on the horizontal axis should be understood as $\langle z\rangle$.
    }
    \label{fig:comparison}
\end{figure*}

In order to constrain $\Omega_R+3\Omega_Q$ we must reconstruct $D$ and $H_\mathcal{D}$ observationally {\em without} invoking assumptions based on the $\Lambda$CDM model or the entire FLRW class of models. We can do this following the bootstrap-based symbolic regression procedure recently introduced in \cite{PRD}. Indeed, we will here use the reconstructed $\langle D_A\rangle$, $H_\mathcal{D}$ and their derivatives obtained for the work presented in \cite{PRD}, to here obtain the constraints on $\Omega_R + 3\Omega_Q$. The reconstructions of \cite{PRD} were obtained by using supernova data from Pantheon+ \cite{pantheon} and baryon acoustic oscillations (BAO) data from BOSS, eBOSS and DESI \cite{bao1, bao2, bao3, bao4, bao5, bao6, bao7, bao8}. Details on the bootstrap-based symbolic regression method can be found in \cite{PRD}, and we will here focus on discussing the possibility of constraining $H_\mathcal{D}$ with BAO data. Note first that we only use the radial BAO measurements which constrain the averaged line-of-sight expansion rate, $\mathcal{H}$, multiplied by the sound horizon at the drag epoch, $r_d$ \cite{bao_review}. For a general spacetime, we can write
\begin{align}
    \mathcal{H} = \frac{1}{3}\theta - e^\mu a_\mu + e^\mu e^\nu \sigma_{\mu\nu},
\end{align}
where $\theta$ is the fluid's isotropic expansion, $a^\mu$ its acceleration, $\sigma_{\mu\nu}$ its shear and $e^\mu$ is the line-of-sight. We expect $a^\mu = 0$ on the large scales probed by BAO. We further expect that the shear vanishes on average on sufficiently large scales. If we further assume that the shear field has no preferred direction and that $e^\mu$ is isotropically distributed in the BAO surveys, the last term in $\mathcal{H}$ would also drop out upon averaging. Thus, under fairly broad assumptions we see that the radial BAO indeed can be used to estimate $H_\mathcal{D}r_d$. To summarize, to constrain $H_\mathcal{D}$ directly from the BAO data we merely need to imposing assumptions of large-scale homogeneity and isotropy which is already a requirement for relating spatial averages to $\langle z\rangle$ and $\langle D_A\rangle$ as in equations \ref{eq:z} and \ref{eq:DA}.
\newline\indent
To determine $H_\mathcal{D}$ we must further assume a value for $r_d$. Usually, one would use the $\Lambda$CDM Planck value, but we here marginalize over $r_d$ in an interval around this value to minimize $\Lambda$CDM model-dependence. Specifically, we marginalize over $\pm 2$Mpc around the Planck best-fit value. This is sufficient to reduce sensitivity to the exact Planck $\Lambda$CDM best-fit value while retaining the assumption that standard early-universe physics provides a good approximation for the sound horizon scale.
\newline\newline
\begin{figure*}
    \centering
    \includegraphics[width=0.65\columnwidth]{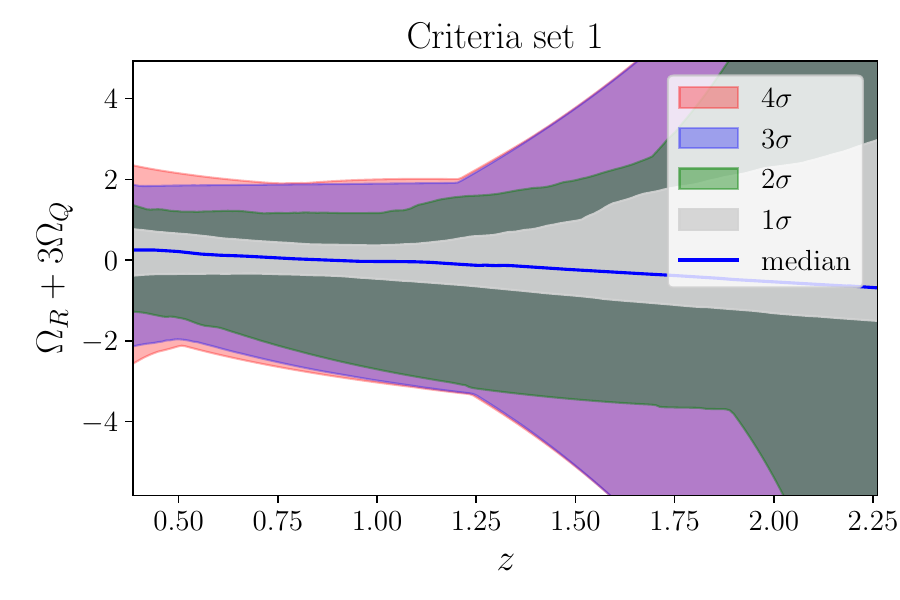}
    \includegraphics[width=0.65\columnwidth]{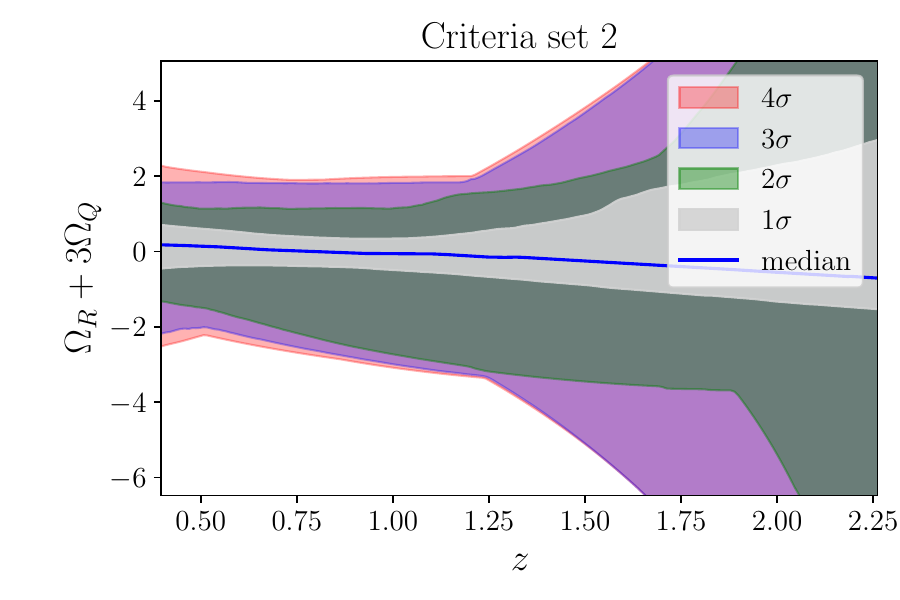}
    \includegraphics[width=0.65\columnwidth]{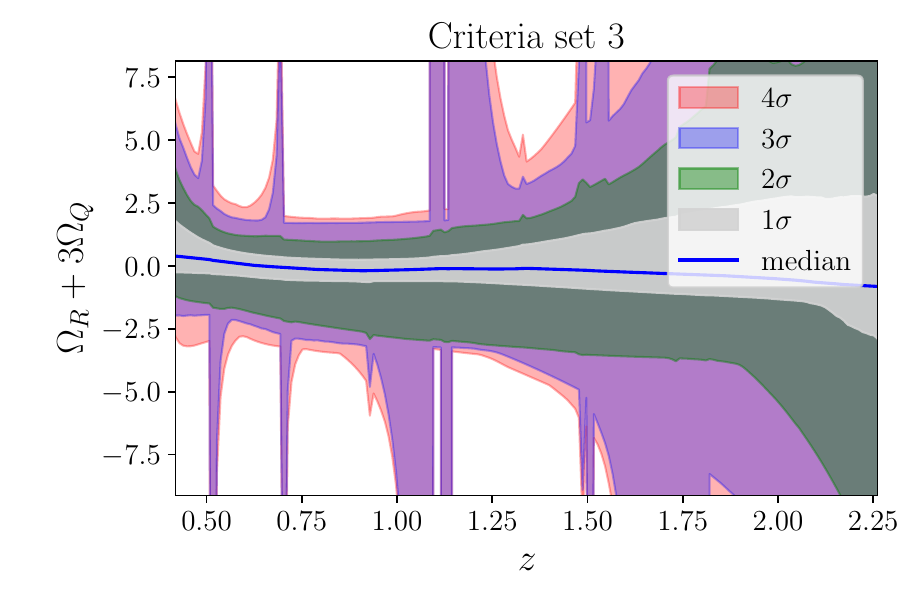}\\
    \includegraphics[width=0.65\columnwidth]{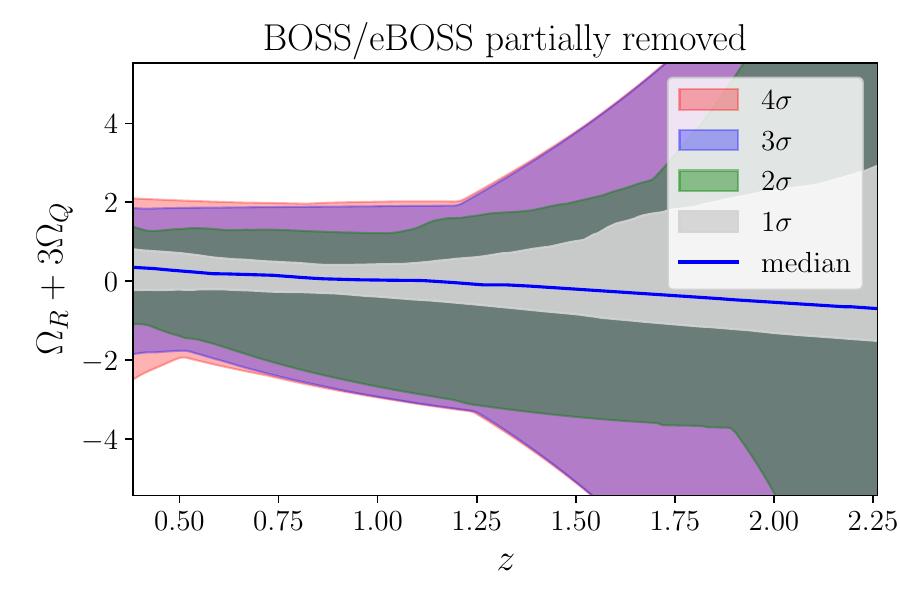}
    \includegraphics[width=0.65\columnwidth]{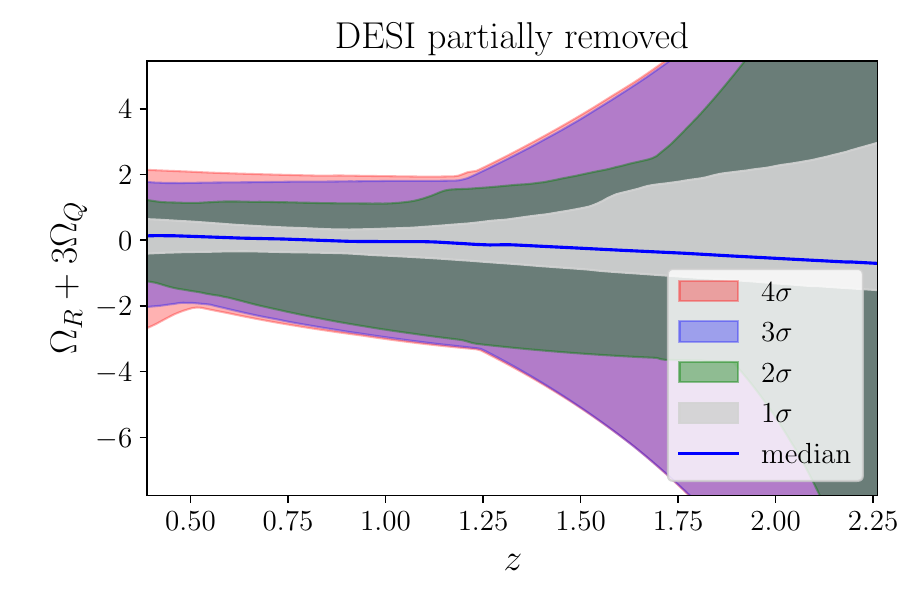}
    \caption{Constraints on $\Omega_R+3\Omega_Q$ in terms of median and percentiles that for Gaussian distribution would correspond to $1,2,3,4\cdot\sigma$.  The results were obtained from reconstructing $\langle D_A\rangle $ and $H_\mathcal{D}$ as described in \cite{PRD}. The redshift $z$ on the horizontal axis should be understood as $\langle z\rangle$.
    }
    \label{fig:results}
\end{figure*}
Figure \ref{fig:results} shows five different versions of the resulting constraints on $\Omega_R + 3\Omega_Q$. These different constraint versions were obtained using slightly different BAO data combinations and different approaches for selecting symbolic expressions reconstructing the BAO data. The details on the different data sets and selection procedures are as in \cite{PRD}. Here, we will briefly note that constraints obtained under the captions ``criteria set i'', $i = 1,2,3$ were obtained using different selection criteria and using DESI data release 2 together with a single eBOSS data point at low redshift to constrain $H_\mathcal{D}$ and $H_\mathcal{D}'$. The two other constraints were obtained using different combinations of DESI data release 1 and (e)BOSS data. 
\newline\indent
The quantitative results shift somewhat when using different data sets, emphasizing the significance of hyper parameter choices for the symbolic regression methods as well as the significance of employing different data combinations. Importantly, although bootstrap-based symbolic regression does not assume a cosmological model, there is still technical model dependence since the results will depend on the choice of hyper parameters of the individual symbolic regression algorithms, the algorithms themselves and expression-selection criteria. The differences seen in the constraints in the five subfigures suggest that these choices are not negligible. In addition, the FLRW test cases of \cite{PRD} indicate that the median value should not be expected to represent the true values. Instead, it is the uncertainty bands that one should mainly consider (as is indeed always the case). Thus, while the mean values are indicated in the plots, the robustness of the method mainly lies in identifying uncertainty bands containing the true values. This is emphasized in figure \ref{fig:comparison} which shows a comparison of the median, $1\sigma$ and $2\sigma$ bands from the five different constraint versions. The comparison shows that the medians overall follow the same trend and the $1-2\sigma$ bands are almost indistinguishable except for those obtained using criteria 3 which are slightly different. Note here, that the constraints shown under the caption ``Criteria set 3'' are based on selection criteria that did not reject symbolic expressions that were undefined or divergent in the studied redshift interval (this is the reason for the spikes in the uncertainty bands).
\newline\indent
The flat FLRW expectation of $\Omega_R + 3\Omega_Q = 0$ lies within one standard deviation of the median. Although this is an important FLRW consistency check, the uncertainty bands are so broad that it is not possible to exclude significant cosmic backreaction with the current constraints; considering equation \ref{eq:one} it is clear that if $\Omega_R + 3\Omega_Q$ is as small as of, say, order $\sim 0.01$, this would mean that backreaction is affecting the large-scale dynamics of the Universe at percent level. This situation would be important in high-precision cosmology, and can clearly not be excluded with the current constraints.

\section{Summary, discussion and conclusion}
We have applied the recently introduced method of \cite{preprint} to place observational constraints on the backreaction parameter combination $\Omega_R + 3\Omega_Q$. The resulting bounds are consistent with a flat FLRW cosmology and thus with the standard $\Lambda$CDM model at the one‑sigma level. However, since we are constraining $\Omega_R+3\Omega_Q$ and not the two components individually, it is important to note that this specific combination can vanish/be very small, without the dynamically important combination $\Omega_r+\Omega_Q$ being so (see e.g. section 3.3.2 of \cite{buchert} for an example). Additionally, the uncertainty bands of the constraints remain substantial, implying that non‑negligible backreaction effects in our universe cannot presently be ruled out. The large uncertainties are due to the low number of BAO data points, marginalizing over $r_d$ and to the constraints being very conservative as we have only implemented very few modeling assumptions such as through hyperparameter choices for symbolic regression algorithms and by assuming that the Dyer-Roeder approximation can be neglected. Tighter constraints could in principle be obtained by imposing specific modeling assumptions on $R_\mathcal{D}$ and $Q_\mathcal{D}$. For example, \cite{preprint} analytically explored scaling solutions corresponding to the assumption $Q_\mathcal{D}\propto a_\mathcal{D}^{-n}$. We have deliberately refrained from imposing such constraints here, as there is currently no clear physical justification for expecting this functional form of $Q_\mathcal{D}$. In the absence of well-motivated backreaction models, we consider it more appropriate to focus on constraining backreaction effects introducing as few model assumptions as possible.
\newline\indent
In a complementary approach, \cite{soon} derived the first direct observational constraints on $\Omega_{R,0}$ and $\Omega_{Q,0}$ in the local universe through reconstructions based on CosmicFlows‑4++ data \cite{flows}, finding a small (sub-percent) value for $\Omega_{Q,0}$ but significant curvature in our local neighborhood. While the method introduced in \cite{preprint} and employed here does not allow $\Omega_R$ and $\Omega_Q$ to be constrained separately, it instead enables constraints over an extended redshift range and on hypersurfaces that are directly related to the observed redshift–distance relation. In contrast, the backreaction constraints from \cite{soon} specifically relate to the hypersurfaces comoving with the matter contained within the region. The region is known to have significant bulk flow (see e.g. \cite{bulk, bulk2}), implying that the region comoving with the fluid deforms significantly with time. The regions probed by the Pantheon+ (at $z\geq0.38$ considered here) and BAO data considered here are expected to be so large that such peculiar flow is not important. The two approaches presented in \cite{preprint} and \cite{soon} thus probe different regimes and provide complementary information.
\newline\indent
Finally, we emphasize that although the constraints presented here are qualitatively robust, they are necessarily exploratory since they are based on a newly proposed bootstrap-based symbolic regression method \cite{PRD}. This is emphasized by the quantitative shifts in the $3-4\sigma$ uncertainty bands when considering different data combinations and applying different symbolic expression selection criteria. The results should therefore not be over-interpreted, but instead viewed as a first step toward more robust and precise future constraints on cosmic backreaction based on analyses of cosmological data using minimal assumptions. Furthermore, the obtained constraints are too weak to be genuinely interesting beyond a proof-of-principle and current status demonstration. With the large amounts of cosmological data expected within the next decade, it is expected that significantly tighter constraints will become possible.

\begin{acknowledgments}
The author thanks Asta Heinesen for correspondence regarding the preprint \cite{preprint}, Thomas Buchert for pointing out that $\Omega_r + 3\Omega_Q$ may vanish while $\Omega_R + \Omega_Q$ remains dynamically important, and the anonymous referee for comments that significantly helped improve the presentation of the results. The author is funded by Villum Fonden under grant VIL53032.
\newline\newline
The expansion rate and angular diameter distance reconstructions used in this work were obtained using the following algorithms: AIFeynman \cite{feynman1, feynman2}, QLattice \cite{lattice}, ITEA \cite{itea}, GeneticEngine \cite{genetic} (all through cp3-bench \cite{bench}), and PySR \cite{pysr}.

\end{acknowledgments}

\end{document}